\documentclass[12pt]{iopart}
\usepackage{graphicx}
\usepackage{amssymb}

\begin{document}

\title{Few Electron Limit of n-type Metal Oxide Semiconductor Single Electron Transistors}  

\author{Enrico Prati$^1$, Marco De Michielis$^1$, Matteo Belli$^1$, Simone Cocco$^1$, Marco Fanciulli$^{1,2}$, 
Dharmraj Kotekar-Patil$^3$, Matthias Ruoff$^3$, Dieter P. Kern$^3$, David A. Wharam$^3$, 
Jan Verduijn$^{4,5}$, Giuseppe C. Tettamanzi$^{4,5}$, Sven Rogge$^{4,5}$, 
Benoit Roche$^6$, Romain Wacquez$^{6,7}$, Xavier Jehl$^6$, Maud Vinet$^7$, Marc Sanquer$^6$}
\address{$^1$ Laboratorio MDM, CNR-IMM, Via Olivetti 2, I-20864 Agrate Brianza, Italy}
\address{$^2$ Dipartimento di Scienza dei Materiali, Universit\`a degli Studi Milano-Bicocca, Via Cozzi 53, I-20125 Milano, Italy}
\address{$^3$ Institute of Applied Physics, Universit\"{a}t T\"{u}bingen, Auf der Morgenstelle 10, 72076 T\"{u}bingen, Germany }
\address{$^4$ Kavli Institute of Nanoscience, Delft University of Technology, Lorentzweg 1, 2628 CJ Delft, The Netherlands}
\address{$^5$ Centre for Quantum Computation and Communication Technology, School of Physics, University of New South Wales, Sydney, New South Wales 2052, Australia}
\address{$^6$ Service de Physique Statistique, Magn\'{e}tisme et Supraconductivit\'{e}, Institut Nanosciences et Cryog\'{e}nie, Commissariat \`{a} l'\'{E}nergie Atomique Grenoble and Universit\'{e} Joseph Fourier, F-38054 Grenoble, France}
\address{$^7$ CEA-LETI Minatec, Grenoble, F-38054, France}
\eads{\mailto{enrico.prati@cnr.it}, \mailto{marc.sanquer@cea.fr}}

\begin{abstract}
We report electronic transport on n-type silicon Single Electron Transistors
(SETs) fabricated in Complementary Metal Oxide Semiconductor (CMOS) technology.
The n-MOSSETs are built within a pre-industrial Fully Depleted Silicon On Insulator (FDSOI) technology with a silicon thickness down to 10 nm on 200 mm wafers. The nominal channel size of 20$\times$20 nm$^{2}$ is obtained by employing 
electron beam lithography for active and gate levels patterning. The Coulomb blockade stability diagram is precisely resolved at 4.2 K and it exhibits large addition energies of tens of meV. The confinement of the electrons in the quantum dot has been modeled by using a Current Spin Density Functional Theory (CS-DFT) method.
CMOS technology enables massive production of SETs for ultimate nanoelectronics and quantum variables based devices.
\end{abstract}

\submitto{NT}
\maketitle

\section{Introduction}

The fabrication of semiconductor quantum dots in CMOS technology provides the ground for integration 
of existing microelectronics with ultimate nanoelectronics for quantum circuits.
We report on the realization of Single Electron Transistors (SETs) in pre-industrial CMOS technology at the limiting channel size of 20$\times$20 nm$^2$ and on the characterization of the quantum transport at cryogenic temperature, enabling the detection of the first electron in the SET.


Silicon SETs based on CMOS technology represent a
natural environment for realizing scalable electron spin and
orbital quantum devices for quantum information processing 
\cite{Friesen03} because of long coherence times \cite{Nature11} and
scalability.
The need to create a workable Hilbert space with good quantum numbers 
\cite{Friesen07} at increasingly high temperatures towards room temperature operability implies the reduction of the size of the quantum dots down to the current limits of fabrication. \cite{Pauliac-Vaujour_JJAP2011}
The drawback is an increasing sensitivity of the confinement potential 
imposed by the control gate to impurities and roughness, and a consequent impact on the
electronic shell structure of the quantum dot.
In the past, some approaches for creating single electron silicon quantum dots have been explored, including the
employment of gate induced two dimensional electron gas (2DEG) at the Si/SiO$_2$ interfaces \cite{Xiao10}, 
global gate controlled edge roughness of an ultra small nanowire \cite{Hiramoto_JJAP2008},
local gate equipped underlap geometry where the gate is not at the state-of-the-art and consequently enforces the use of large gate voltages \cite{Shin_APL2010}, triple-layer gate stacks \cite{Lim_Nanotechnology2011} and Si/SiGe modulation-doped heterostructures \cite{Simmons07}.

Our approach consists in confining electrons in a well defined by lateral doping modulation. Indeed barriers are consequence of undoped silicon below spacers which are on both sides of the gate. \cite{Hofheinz_APL2006} In the past, this approach has been exploited to fabricate compact double and triple quantum dots realized with only two gates. \cite{Pierre09}
The two main advantages are on one hand its compactness, as regard to state-of-the-art previous works on Si including additional upper gate \cite{Liu_PhysRevB2008}, and on the other one the detection of a clear single electron regime.

The use of CMOS technology in quantum dot fabrication takes benefit from its reproducibility and reliability, as well as the co-integration of quantum circuits with traditional CMOS.

In Section \ref{Sec:Fabrication} the fabrication of the CMOS device is illustrated while in  Section \ref{Sec:Simulation} the simulation of the electron confinement is discussed.
In Section \ref{Sec:Characterization} the quantum transport obtained in the 20$\times$20 nm$^{2}$ samples at the temperature $T$= 4.2 K is presented.


\section{Fabrication of the devices}
\label{Sec:Fabrication}

For a repeatability and reliability purpose, our MOSSETs are built within a pre-industrial Fully Depleted Silicon On Insulator (FDSOI) technology on 200 mm wafers. Only a few modules of the device are slightly modified, such as the gate stack and the source/drain (S/D) implantation. Since we aim to scale our MOSSET down to 20 nm in both gate length and gate width dimensions, electron beam lithography was used for active and gate levels patterning. That results with dots as small as 20$\times$20$\times$10 nm$^{3}$.

\begin{figure}[t]
\begin{center}
\includegraphics[width=0.6\textwidth]{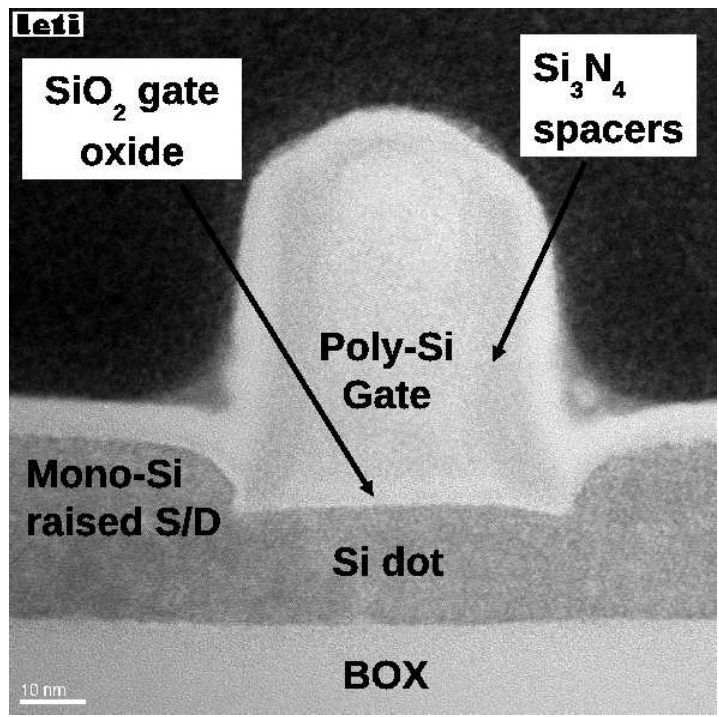}
\end{center}
\begin{center}
\large{(a)}
\end{center}
\begin{center}
\includegraphics[width=0.6\textwidth]{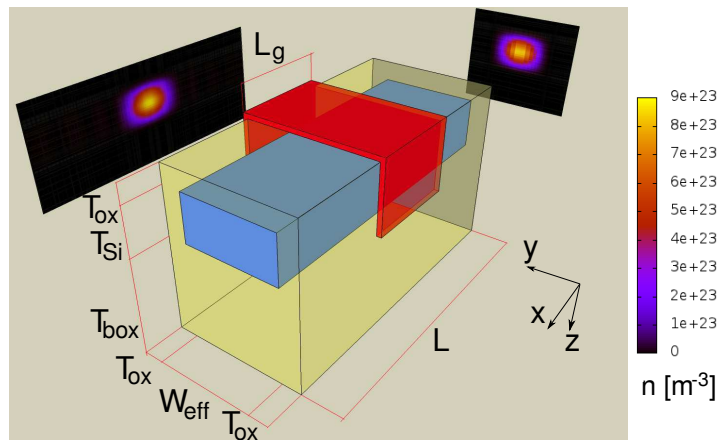}
\end{center}
\begin{center}
\large{(b)}
\end{center}
\caption{a) TEM micrograph of a typical scaled MOSSET adapted from the FDSOI technology. The polysilicon gate length is 22 nm long, and dot thickness is 17 nm (dot width is not shown on this view). The SiO$_{2}$ gate oxide of 4 nm, the nitride spacers of 11 nm and the raised source/drain are also clearly visible. b) Scheme of the simulated device. Light-blue represents the silicon nanowire. The tri-sided gate is highlighted in red whereas the source and drain contacts are in blue and silicon dioxide is in yellow. 
The electron density in the central planes of the nanowire is reported as color-scaled projections for the case of single electron occupancy of the dot.} 
\label{Fig:device_picture+scheme}
\end{figure}

The undoped SOI layer is thermally thinned down to reach a silicon thickness of either 13 nm or 20 nm, depending on the sacrificial oxide thickness.
After a first e-beam lithography, the SOI layer is etched to pattern the active area above the buried oxide (BOX). 
As a result of these first process steps a silicon nanowire is obtained.
This mesa isolation allows to have wrapping gates on three sides of the nanowires.
The silicon nanowire is then thermally oxidized on top and sides, resulting in a 4 nm thick SiO$_2$ formation and a final silicon thickness $T_{Si}$=10 nm (17 nm) prior to polycrystalline silicon deposition. 
 A second e-beam lithography is performed to define the tri-sided gate. Self-aligned silicon nitride spacers are formed on both the source- and drain-side of the gate to protect the underlying silicon from the subsequent doping steps. Epitaxy is then performed to raise the source and drain, and finally arsenic is implanted at high dose (leading to a typical concentration above 10$^{20}$ cm$^{-3}$) to create metallic-like S/D contacts. The resulting junction profile is such that the device is non-overlapped, i.e. the transistor's channel is separated from the source and drain by the small low-doped region below the spacers. This non-overlapped geometry is responsible for the SET behavior \cite{Hofheinz_APL2006}. Figure \ref{Fig:device_picture+scheme}a shows a transmission electron micrograph of a tri-sided gate nanowire MOSSET coming from a wafer used for the morphological characterization with channel thickness $T_{Si}$=17 nm, oxide thickness $T_{ox}$=4 nm and gate length $L_g$=22 nm realized with a process identical to the samples reported throughout the text.

\section{Simulation of the electrostatic confinement}
\label{Sec:Simulation}

In order to evaluate the confinement of the electrons induced by the applied electrostatic potential we used a three dimensional self-consistent simulator based on Current Spin Density Functional Theory (CS-DFT) in the framework of the NanoTCAD ViDES package. \cite{Lisieri_JCE2007, NanoTCAD_VIDES, Fiori_Trans_Nanotech_2005, Fiori_Trans_Nanotech_2007} 
It solves the many particle Schrödinger equation by means of CS-DFT, by transforming the many electron problem in a single-electron problem with exchange-correlation potential. It embraces the local density approximation and the effective mass approximation with parabolic bands. The method provides the ground state of the system for each occupation number $N$ of electrons. They fill the lowest single-particle Kohn-Sham eigenvalues calculated for each spin and each pair $j=1,2,3$ of the three $\Delta$ valley pairs aligned to the main directions in the $k$ space. 

The final shape of the confining potential energy is reached by self-consistently solving the Poisson equation. The effective nature of the confinement is due to the combination of the band alignment between the silicon and silicon dioxide and of the applied external potentials. 

The SET is modeled as a silicon nanowire with length $L$ along the x direction and with a rectangular section of $W_{eff}$$\times$$T_{Si}$ area on the y-z plane with source and drain contacts at the head and tail of the nanowire on the top of a silicon dioxide slab of thickness $T_{box}$, see Figure \ref{Fig:device_picture+scheme}b. The device has a tri-sided gate structure and the gate is insulated from the nanowire with a $T_{ox}$ thick silicon dioxide layer.

The many particle problem is solved in the entire silicon nanowire and in a 1 nm-thick region inside the surrounding oxide. Potentials to the tri-sided gate, source and drain regions are applied whereas a zero electric field z-component boundary condition is forced at the bottom of the device. 
We simulate the electrostatic behavior of the SETs imposing an effective gate length $L_{g,eff}$= 22 nm, an effective width $W_{eff}$=10 nm to take into account the variability of the lithography process on the nominal length $L_g$=20 nm and both the lithography process variability and the oxidation reduction on the nominal width $W$=20 nm.

In Figure \ref{Fig:device_picture+scheme}b the electron density in the central planes of the nanowire is reported as color-scaled planes, showing a strong confinement in the region under the gate contact biased at $V_{g}$=26 mV when one electron is in the dot.
The first addition energy $E_{(2,1)}$ which provides the energy separation between the one electron state and the two electrons state has been calculated. By using the effective size of $L_{g,eff}$=22 nm, $W_{eff}$=10 nm we determined an addition energy of $E_{(2,1)}$=17 meV. 
Note that our simulations take into account all the couplings of the SET
with the gate electrode but also with the source and drain contacts.
Our approach does not take into account the possible disorder caused by the impurities and the interface roughness, which are expected to further enhance the confinement of the wavefunction in such small devices.


\section{Quantum transport experimental results}
\label{Sec:Characterization}

The investigation of the electronic transport through the devices yields both the electron
filling as a function of the QD gate voltage $V_g$, as well
as the addition energies $E_{(n+1,n)}$ needed to add the $(n+1)^{th}$
electron when $n$ electrons occupy the quantum dot. The characterization at the
temperature $T$=4.2 K of several nominally identical samples like those
described in Section \ref{Sec:Fabrication} demonstrate that the transistors operate as
single electron devices, with relatively high addition energies.

\begin{figure}[t]
\begin{center}
\includegraphics[width=0.5\textwidth]{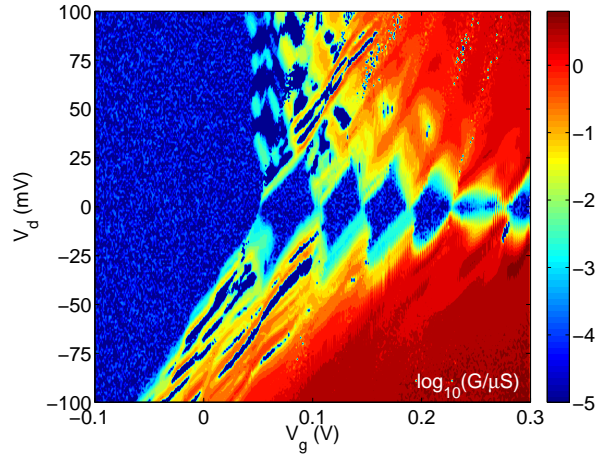}
\end{center}
\begin{center}
\large{(a)}
\end{center}
\begin{center}
\includegraphics[width=0.5\textwidth]{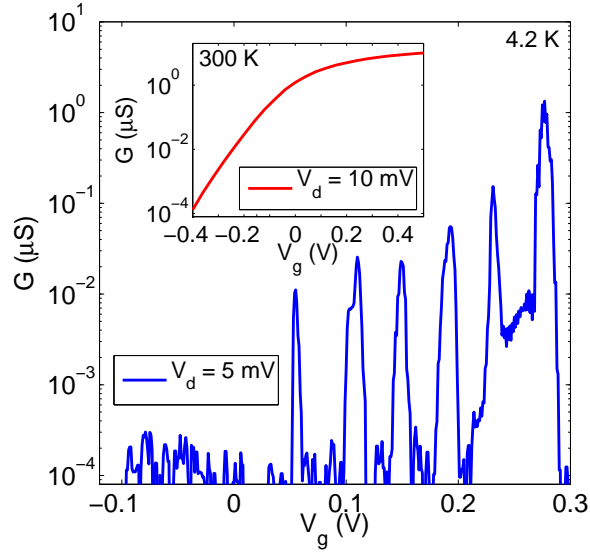}
\end{center}
\begin{center}
\large{(b)}
\end{center}
\caption{a) Differential conductance as a function of $V_{d}$ and $V_{g}$ for a $L_{g}$$\times$$W$=20$\times$20 nm$^2$
and b) conductance at $T$=4.2 K and at $T$=300 K (inset) as a function of $V_{g}$.
} 
\label{Fig:stability_diagrams}
\end{figure}

In Figure \ref{Fig:stability_diagrams}a the stability
diagram of a typical device with of $L_g=20$ nm and $W=20$ nm is shown.
The coupling with the source, drain and gate are $C_s= 1.0$ aF, $C_d= 1.3$ aF and $C_g= 4.0$ aF respectively. 
The lever arm factor, which allows to convert the voltage spacing between the peaks into addition energies and which is defined as $\alpha=C_g/(C_g+C_s+C_d)$, assumes a value of 0.64.
The first addition energy of the device shown in Figure \ref{Fig:stability_diagrams}a is $E_{(2,1)}$= 30 meV.
Figure \ref{Fig:stability_diagrams}b shows the Coulomb blockade observed from $N$=0 up to a filling number
of 6 in the range from -0.1 V to 0.3 V. In the inset the results of the conductance measurement performed at room temperature are reported. 

In these small devices it is possible to reduce the electron occupancy of the quantum dot
down to the empty state, hence the transition from $1$ to $0$ electron occupancy of the SET becomes observable, a key step towards single electron charge and spin manipulation.

The experimental confirmation is obtained by the lack of Coulomb diamond and any detectable current (up to $V_d$=100 mV) below the first reported Coulomb oscillation at $V_g$=40 mV. Moreover we notice that there is no significant shift between the threshold voltage obtained at $T$=300 K ($\sim$ -40 mV calculated as the maximum of the second derivative of the drain current with respect to the gate voltage, see inset of Figure \ref{Fig:stability_diagrams}b) and the gate voltage position of the first Coulomb peak at low temperature ($V_g$= 40 mV), taking into account the thermal broadening at $T$=300 K. 
 
The presented samples are smaller than similar samples previously studied ($L$,$W$=30-40 nm, \cite{Hofheinz_APL2006} and \cite{Roche_APL2012}). The silicon thickness ($T_{Si}$=10 nm) and the spacer length (11 nm) have been scaled down accordingly to obtain a good trade-off between the first electron orbital-electrodes coupling and a large Coulomb energy. By contrast in previously reported similar devices \cite{Hofheinz_APL2006} the first electrons in the accumulation channel are not detected and the first measurable current is at significantly larger gate voltage than the threshold voltage at room temperature. 
On the contrary for similar samples but without nitride spacers \cite{Lim_Nanotechnology2011} - and therefore smaller effective channel length - the first electrons are on shallow donors in the body of the silicon and detected at gate voltage much below the threshold voltage at room temperature. \cite{Pierre_NatureNanotech2012,Leti_APL2011} In that case there is no MOSSET effect in the accumulation channel. 

Similar results have been obtained in nominally identical devices. Because of the residual disorder determined by the interface roughness, the electron wavefunction is further confined so the first addition energies are even larger than those predicted from the simulation of the devices and are subject to some variability.
From the addition energies found in typical devices, an effective radius of less than 12-15 nm can be estimated.

\section{Conclusions}
Our results show that CMOS technology used allows reliable fabrication of 
single electron transistors with gate size of 20$\times$20 nm$^{2}$.
Strong confinement has been predicted by CSDFT simulations and a clearly resolved Coulomb blockade pattern at 4.2 K with high addition energies of more than 20 meV exceeds the simulated values.
The fabrication of SETs in CMOS technology provides the possible ground for integration between traditional microelectronics and quantum circuit oriented nanoelectronics.

\ack
The research leading to these results has received funding from the
European Community's seventh Framework (FP7 2007/2013) under the 
Grant Agreement No. 214989. The samples subject of this work 
have been designed and made by the AFSID Project Partners,
http://www.afsid.eu.
\\
We acknowledges Prof. Giuseppe Iannaccone and Prof. Gianluca Fiori for providing the NanoTCAD ViDES simulator and for fruitful discussions. 
\\
M. F. thanks support by the ELIOS project granted by Cariplo Foundation.
\\
S. R. thanks support by an ARC-Future Fellowship project ID: FT100100589. 
\\
G. C. T. thanks support by an ARC-DECRA Fellowship project ID: DE120100702






\section*{References}
\bibliographystyle{unsrt.bst}
\bibliography{biblio.bib}

\end{document}